\documentclass[twocolumn,showpacs,preprintnumbers,amsmath,amssymb,prb,superscriptaddress]{revtex4-1}
\usepackage{multirow}
\usepackage{amsfonts}
\usepackage[english]{babel}
\usepackage[T1]{fontenc}
\usepackage{times}
\usepackage{mathrsfs}
\usepackage{graphicx}
\usepackage{dcolumn}
\usepackage{bm}
\usepackage[colorlinks,bookmarks=true,citecolor=blue,linkcolor=red,urlcolor=blue]{hyperref}
\usepackage[tight, FIGTOPCAP, hang, raggedright, nooneline]{subfigure}

\newcommand{\bb}[1]{\mathbf{#1}}

\begin{document}

\title{Characterization of quasiholes in fractional Chern insulators}

\author{Zhao Liu}
\affiliation{Department of Electrical Engineering, Princeton University, Princeton, New Jersey 08544, USA}
\author{R. N. Bhatt}
\affiliation{Department of Electrical Engineering, Princeton University, Princeton, New Jersey 08544, USA}
\author{Nicolas Regnault}
\affiliation{Department of Physics, Princeton University, Princeton, New Jersey 08544, USA}
\affiliation{Laboratoire Pierre Aigrain, ENS-CNRS UMR 8551, Universites P. et M. Curie and Paris-Diderot, 24, rue Lhomond, 75231 Paris Cedex 05, France}

\date{\today}

\begin{abstract}
We provide a detailed study of the Abelian quasiholes of $\nu=1/2$ bosonic fractional quantum Hall states on the torus geometry and in fractional Chern insulators. We find that the density distribution of a quasihole in a fractional Chern insulator can be related to that of the corresponding fractional quantum Hall state by choosing an appropriate length unit on the lattice. This length unit only depends on the lattice model that hosts the fractional Chern insulator. Therefore, the quasihole size in a fractional Chern insulator can be predicted for any lattice model once the quasihole size of the corresponding fractional quantum Hall state is known. We discuss the effect of the lattice embedding on the quasihole size. We also perform the braiding of quasiholes for fractional Chern insulator models to probe the fractional statistics of these excitations. The numerical values of the braiding phases accurately match the theoretical predictions.
\end{abstract}

\maketitle

\section{Introduction}

One of the most remarkable features of topological ordered phases of matter is the emergence of fractionally charged excitations. These anyons may obey fractional or even non-Abelian statistics.\cite{leinaas-77ncb1,Arovas-PhysRevLett.53.722} This unique property is the key to implement topological quantum computation.\cite{Nayak-RevModPhys.80.1083} Among the physical systems that exhibit anyonic excitation, the fractional quantum Hall (FQH) effect is the most prominent example. There, an electron gas confined in two dimensions and penetrated by a uniform strong magnetic field can host anyonic excitations.\cite{Laughlin:1983p301,Moore:1991p165} Both theoretical \cite{Bonderson-PhysRevLett.96.016803,Bishara-PhysRevB.80.155303,Wan-PhysRevB.77.165316,Toke-PhysRevLett.98.036806,Prodan-PhysRevB.80.115121,Storni-PhysRevB.83.195306,Johri-PhysRevB.89.115124,Wu-PhysRevLett.113.116801} and experimental \cite{Camino-PhysRevB.72.075342,Camino-PhysRevLett.98.076805,dolev-08n829,Venkatachalam-nat09680} characterization of anyons in FQH systems has been attracting a great deal of interest in current condensed matter research.

Recently, variants of the FQH effect on two-dimensional (2D) lattices have received a large interest. In particular, this includes the Chern bands or Chern insulators (CIs) that are the analog of a single Landau level.\cite{Tang-PhysRevLett.106.236802,Sun-PhysRevLett.106.236803,neupert-PhysRevLett.106.236804} Strong interaction leads to correlated phases referred to as fractional Chern insulators (FCIs).\cite{Parameswaran2013816,BERGHOLTZ-JModPhysB2013,Neupert-2014arXiv1410.5828N} These phases exhibit similar features to the FQH cousins. Some of the FCIs are direct lattice generalizations of the FQH effect,\cite{Kol:1993p82,Hafezi:2007p67,Sorensen:2005p58,Moller:2009p184,Palmer:2006p63,Kapit-PhysRevLett.105.215303,PhysRevLett.111.186804} similar to the well-known Hofstadter model. In other cases, the absence of net external magnetic field is a major advantage.\cite{neupert-PhysRevLett.106.236804,sheng-natcommun.2.389,regnault-PhysRevX.1.021014,Wu-PhysRevB.85.075116,Wang-PhysRevLett.108.126805,Lauchli-PhysRevLett.111.126802,Liu-PhysRevLett.109.186805,Yao-PhysRevLett.110.185302} FCIs are usually based on tight-binding models with short-range repulsion, a natural set-up in electronic materials. But other models far from the tight-binding limit have been proposed in the context of ultracold gases such as optical flux lattices.\cite{ofl,cooperdalibard,Sterdyniak-2014arXiv1410.0357S}

The study of the excitations in FCIs via the energy or the entanglement spectra have been widely used as a way to probe the topological order of these phases. Interestingly, direct characterization of the excitations has been mostly ignored (except for analytic wave functions \cite{Kapit-PhysRevLett.108.066802,2014arXiv1409.3073N}). A basic property such as the spatial extent of the quasiholes is highly relevant for experimental set-ups and measurements. In the case of the FQH effect, this size \cite{Toke-PhysRevLett.98.036806,Prodan-PhysRevB.80.115121,Storni-PhysRevB.83.195306,Johri-PhysRevB.89.115124,Wu-PhysRevLett.113.116801} is a limiting factor for the design interferometers.\cite{Rosenow12:Interferometer} For ultracold gases where one can directly image the density, the spatial extent is required to know if the detection of quasiholes is within the range of the optical resolution.\cite{Douglas-PhysRevA.84.053608}

In this paper, we numerically study the previously ignored quasihole properties in FCIs, namely their spatial extent, charge, and statistics. Motivated by the relevance for the ultracold gas implementation, we focus on strongly interacting bosons and the simplest phase: the Laughlin $\nu=1/2$ phase. We consider various FCI models such as the kagome,\cite{Wang-PhysRevLett.107.146803,Wu-PhysRevB.85.075116} checkerboard,\cite{Sun-PhysRevLett.106.236803,neupert-PhysRevLett.106.236804,regnault-PhysRevX.1.021014} honeycomb,\cite{Haldane-PhysRevLett.61.2015,Wu-PhysRevB.85.075116} and ruby \cite{Hu-PhysRevB.84.155116,Wu-PhysRevB.85.075116} lattices and the Kapit-Mueller model.\cite{Kapit-PhysRevLett.105.215303,Liu-PhysRevB.88.205101} Pinning a single quasihole on the lattice, we propose an empirical relation between the spatial extent of a quasihole in a given FCI model and the one of a quasihole in the corresponding FQH state. This relation simply requires substituting the magnetic length $\ell_B$ in the continuum with an effective lattice-dependent magnetic length $\ell_B^{\textrm{lat}}\equiv\sqrt{A/(2\pi)}$, where $A$ is the area of the lattice unit cell. Since the FQH quasihole can be reliably determined, we thus provide a simple way to predict the spatial extent for any type of quasihole in any FCI model. We also perform the braiding of two quasiholes for the above mentioned FCIs to probe the fractional statistics of these excitations. Despite the moderate system sizes that can be numerically reached by exact diagonalization (up to $24$ unit cells), we accurately recover the predicted statistical phase. Interestingly, the braiding path can be chosen to enclose an integer number of unit cells and thus to cancel any Aharonov-Bohm (AB) contribution to the Berry matrix, thus providing a direct access to the statistical phase.

The paper is organized as follows. In Sec.~\ref{sec:continuum}, we describe the numerical characterization of the Laughlin $\nu=1/2$ FQH quasihole on the torus geometry, including the quasihole size, charge, and statistics. The close relation between this geometry and the FCIs with periodic boundary conditions,\cite{Bernevig-2012PhysRevB.85.075128} exemplifies how the pinning, measurement and braiding of quasiholes are numerically performed. In Sec.~\ref{sec:fci}, we discuss the case of quasihole excitation in FCIs. By studying various models, we provide a simple formula to predict the quasihole size in any FCI model from the quasihole size in the corresponding FQH state by defining a lattice-dependent effective magnetic length. We explicitly show the fractional statistics in the case of Laughlin-like phase in FCIs. Finally, we summarize our results in Sec.~\ref{sec:conclusion}.

\section{FQH Quasiholes on the torus geometry}
\label{sec:continuum}
We consider $N$ bosons of charge $e$ in the lowest Landau level on a
torus spanned by two basic vectors $\textbf{L}_1$ and $\textbf{L}_2$. Assuming the number of flux quanta, $N_s$, through surface of the torus is
an integer, the magnetic translation invariance leads to $\textbf{L}_1\times\textbf{L}_1=2\pi\ell_B^2 N_s$, where $\ell_B$ is the magnetic length. The filling in one Landau level is defined as $\nu\equiv N/N_s$.
We suppose that the bosons interact by the periodic two-body contact Hamiltonian
\begin{eqnarray}
H_{\textrm{int}}&=&\sum_{s,t=-\infty}^{+\infty}\sum_{i<j=1}^N\delta(\bb r_i-\bb r_j+s\bb L_1+t\bb L_2).
\label{fqhHint}
\end{eqnarray}
For this interaction, the two-fold degenerate $\nu=1/2$ bosonic Laughlin state is the densest exact zero-energy ground states.

When we add $N_{\textrm{qh}}$ extra flux quanta, namely $N_s=2N+N_{\textrm{qh}}$, $N_{\textrm{qh}}$ Abelian quasiholes with charge $-e/2$ are nucleated. In the energy spectrum of $H_{\textrm{int}}$, these delocalized quasiholes are associated with a manifold of zero-energy states, whose degeneracy per momentum sector can be predicted using the Haldane's exclusion principle.\cite{Bernevig-2012PhysRevB.85.075128} Each quasihole can be pinned at position $\bb w_k$ by an impurity potential $H_{\textrm{imp}}(\bb w_k)$. A direct diagonalization of $H_{\textrm{int}}+\sum_{k=1}^{N_{\textrm{qh}}}H_{\textrm{imp}}(\bb w_k)$ gives the ground states with $N_{\textrm{qh}}$ localized quasiholes. For $\nu=1/2$ Laughlin states, a $\delta$ impurity potential $H_{\textrm{imp}}(\bb w)=\sum_{i=1}^N\delta(\bb r_i-\bb w)$ can readily keep a quasihole screened and localized at $\bb w$. Thus the states with localized quasiholes preserve zero energy. However, the computational cost of this direct diagonalization is high because localized quasiholes break the translation invariance on the torus and we do not have
good quantum numbers to reduce the many-body Hilbert space dimension. In order to avoid this difficulty, we assume that the gap between the quasihole manifold and other excited states
is much larger than the strength of impurities so impurities cannot mix them. Then we can first diagonalize $H_{\textrm{int}}$ to obtain the quasihole manifold of zero-energy states,
where we have translation invariance. We then diagonalize impurity potentials in the quasihole manifold, whose dimension is much smaller than the one of the full many-body Hilbert space, to obtain the ground states with localized quasiholes. The degeneracy of localized quasihole states on the torus can be predicted by Haldane's exclusion principle as well as conformal field theory.\cite{ardonne-2008-04-P04016}

\begin{figure}
\centerline{\includegraphics[width=\linewidth]{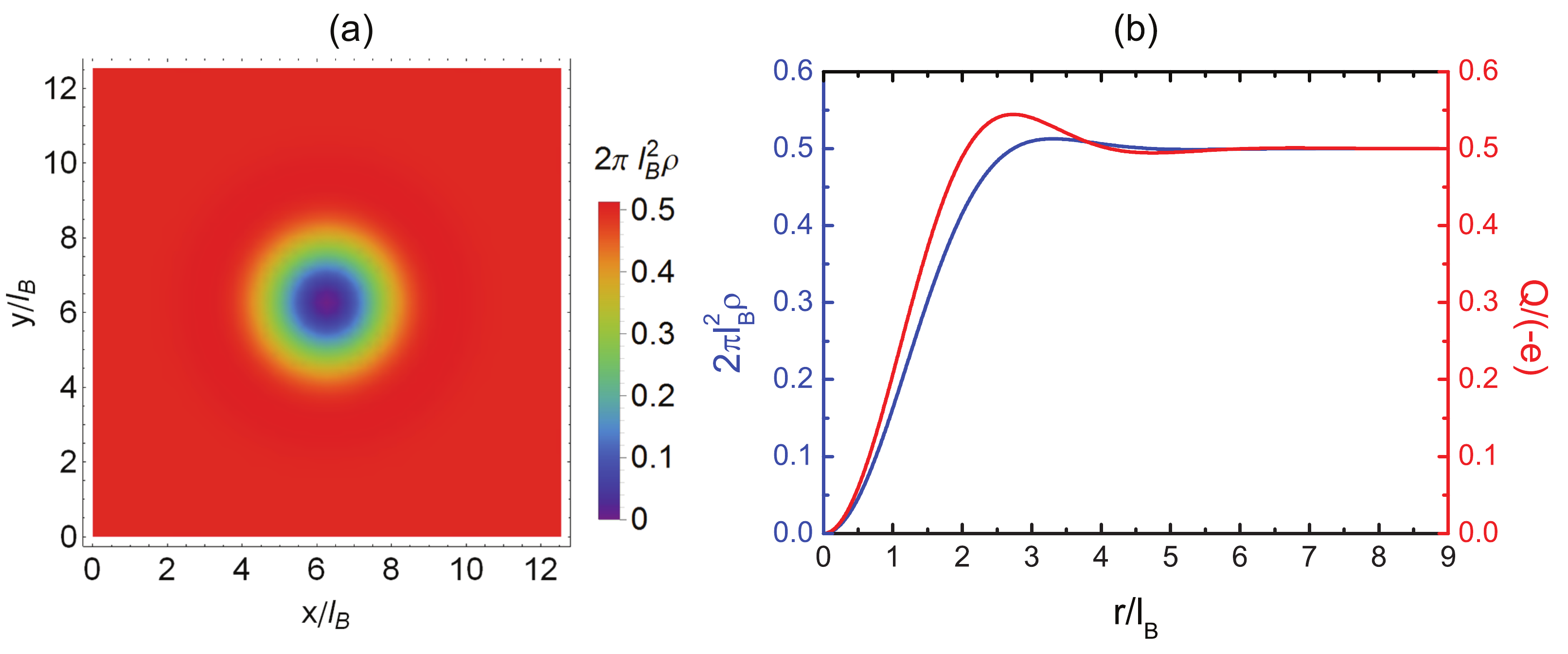}}
\caption{(Color online) (a) The density distribution for the single-quasihole excitation of $\nu=1/2$ bosonic Laughlin states on a square torus with $N=12$ and $N_s=25$.
(b) The radial dependence of the density distribution (blue line) and the excess charge (red line) around the quasihole.}
\label{fqhden}
\end{figure}

\subsection{Single quasihole}
\label{continuuma}
A first characterization of the existence of the quasihole is to compute its spatial extent and charge. For this purpose, we generate a single quasihole by inserting a single flux quantum, namely $N_s=2N+1$. In this case, the quasihole manifold contains $N_s$ zero-energy states. For simplicity, we choose a torus with a square aspect ratio, i.e., $|\textbf{L}_1|=|\textbf{L}_2|$ and $\textbf{L}_1\cdot\textbf{L}_2=0$.
By diagonalizing the impurity in the quasihole manifold of dimension $N_s$, we obtain two zero-energy degenerate ground states.
We could in principle consider any linear combination of these two states as a single pinned quasihole. For the system sizes that we have considered, different choices lead to similar density distribution (with a relative error $\lesssim10^{-4}$ for $N=12$). In order to avoid this arbitrariness while preserving similar results, we compute the average spatial density distribution $\rho$ over these two degenerate states. We find that the quasihole is indeed pinned at the center of the torus where the density is zero [Fig.~\ref{fqhden}(a)].
$\rho$ is isotropic around the quasihole and $2\pi \ell_B^2\rho$ tends to $\nu=1/2$ when the distance $r$ from the quasihole reaches about $5\ell_B$ [in the following we use the numerical value of $\rho(r)$ in the diagonal direction] [Fig.~\ref{fqhden}(b)].
In order to verify the quasihole charge is $-e/2$, we calculate the excess charge defined as
\begin{eqnarray}
Q(r)=2\pi\int_0^{r}[\rho(r')-\rho_0]r' dr',
\label{qhexcesscharge}
\end{eqnarray}
where $\rho(r)$ is the density at the distance $r$ from the quasihole and $\rho_0=\nu/(2\pi\ell_B^2)$ is the uniform density without quasiholes. We find that $Q(r)$ indeed tends to $-e/2$ for $r\gtrsim 6\ell_B$ [Fig.~\ref{fqhden}(b)].

The quasihole radius $R_{\textrm{qh}}^{\textrm{FQH}}$ can be estimated by various methods.\cite{Johri-PhysRevB.89.115124} Here we adopt the definition by the second moment of $\rho(r)$, namely
\begin{eqnarray}
R_{\textrm{qh}}^{\textrm{FQH}}=\sqrt{\frac{\int_0^{r_{\textrm{max}}}|\rho(r)-\rho(r_{\textrm{max}})|r^3 dr}{\int_0^{r_{\textrm{max}}}|\rho(r)-\rho(r_{\textrm{max}})|r dr}},
\label{rqh}
\end{eqnarray}
where $r_{\textrm{max}}$ is the largest distance from the quasihole on our sample. A numerical calculation of Eq.~(\ref{rqh}) shows $R_{\textrm{qh}}^{\textrm{FQH}}\approx1.76\ell_B$. Compared with the sizes of Laughlin quasiholes obtained by the second moment at other fillings ($R_{\textrm{qh}}^{\textrm{FQH}}\approx2.6\ell_B$ at $\nu=1/3$, $R_{\textrm{qh}}^{\textrm{FQH}}\approx3.3\ell_B$ at $\nu=1/4$, and $R_{\textrm{qh}}^{\textrm{FQH}}\approx5.9\ell_B$ at $\nu=1/5$), we notice that $R_{\textrm{qh}}^{\textrm{FQH}}$ is approximately proportional to $1/\nu$.

\subsection{Braiding of two quasiholes}
\label{continuumb}
\begin{figure}
\centerline{\includegraphics[width=\linewidth]{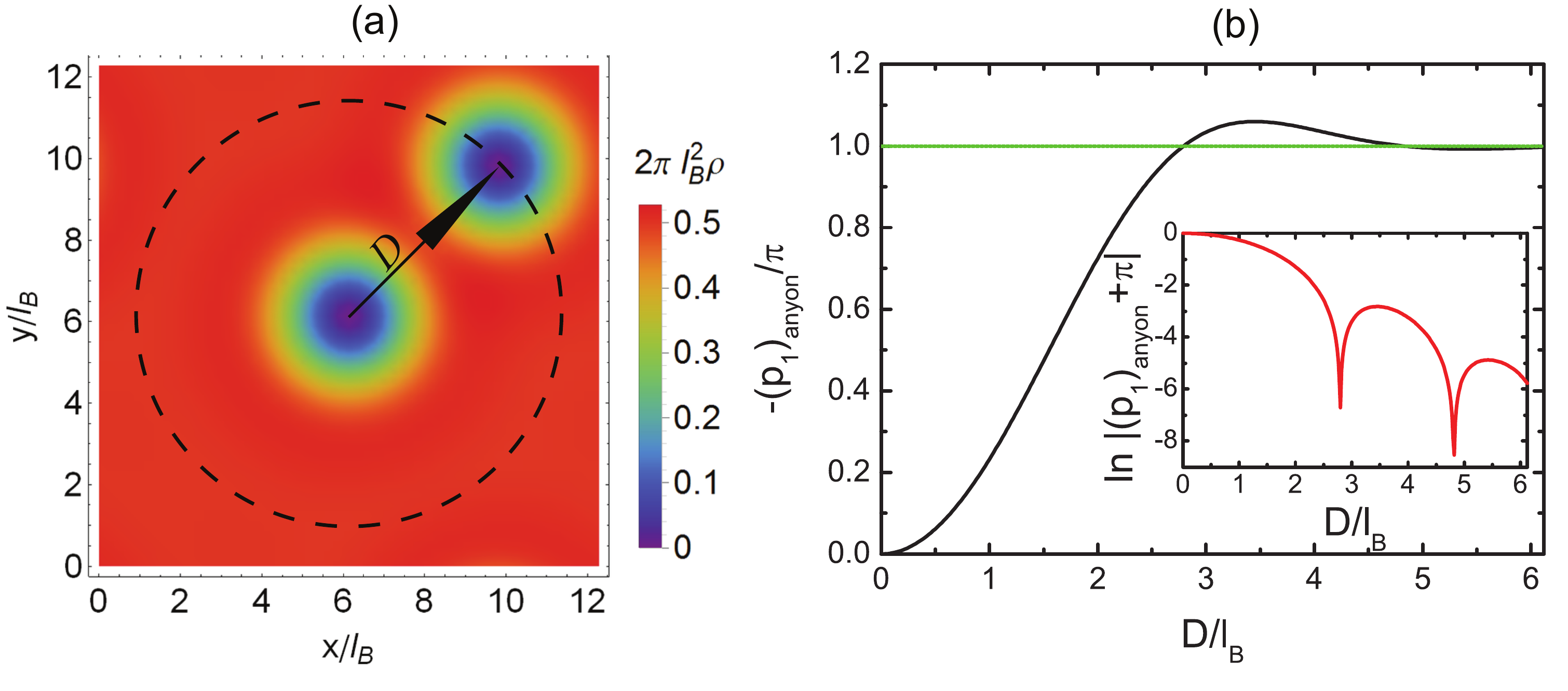}}
\caption{(Color online) (a) The density distribution for the two-quasihole excitation of $\nu=1/2$ bosonic Laughlin states on a square torus with $N=11$ and $N_s=24$. (b) The anyon statistics phase versus the distance between two quasiholes. The inset shows its exponential convergence to the theoretical value (green line). Only $(p_1)_{\textrm{anyon}}$ is shown here hence $(p_1)_{\textrm{anyon}}$ and $(p_2)_{\textrm{anyon}}$ are almost the same.}
\label{fqhbraid}
\end{figure}

Another characterization of the quasiholes is the anyon statistics when we braid them. For simplicity, we generate only two quasiholes by choosing $N_s=2N+2$ and use two $\delta$ impurities to separate and to pin them. Again, by diagonalizing the impurity potentials in the zero-energy quasihole manifold, we obtain two zero-energy degenerate ground states with two localized quasiholes. We compute the average spatial density distribution $\rho$ over these two degenerate states and find that we can indeed pin the quasiholes and separate them [Fig.~\ref{fqhbraid}(a)].

In order to extract the anyon statistics, we move one quasihole clockwise around the other along a circle of radius $D$ [Fig.~\ref{fqhbraid}(a)]. This can be achieved by fixing the position of one impurity and moving the other. The accumulated Berry phase is encoded in the eigenvalues of the unitary Berry matrix
\begin{eqnarray}
\mathcal{B}=\exp\Big\{-2\pi\textrm{i}\int_{0}^{2\pi}\gamma(\theta) d\theta\Big\},
\label{berry}
\end{eqnarray}
where $\theta$ is the polar angle of the mobile quasihole (impurity), $\gamma_{ij}(\theta)=\textrm{i}\langle\psi_i(\theta)|\nabla_{\theta}|\psi_j(\theta)\rangle$ is the Berry connection matrix, and $|\psi_i(\theta)\rangle$ are the degenerate states that we get by diagonalizing the impurity potentials for each $\theta$. By imposing a smooth gauge condition $\langle\psi_i(\theta)|\psi_j(\theta+d\theta)\rangle=\delta_{ij}+\mathcal{O}(d\theta^2)$, we have $\mathcal{B}_{ij}=\langle\psi_i(2\pi)|\psi_j(0)\rangle$. The eigenvalues of $\mathcal{B}$ are $(e^{-\textrm{i}p_1},e^{-\textrm{i}p_2})$, where $p_1$ and $p_2$ are the Berry phases.

The total Berry phase $(p_1,p_2)$ can be split into two parts: one is the AB phase $(p_1,p_2)_{\textrm{AB}}$ caused by moving a single quasihole in the uniform magnetic field along the same path without the other quasihole enclosed; the other comes from the anyonic statistics $(p_1,p_2)_{\textrm{anyon}}$. The AB phase should be $(p_1)_{\textrm{AB}}=(p_2)_{\textrm{AB}}=\pi D^2/2$, and the anyon statistics $(p_i)_{\textrm{anyon}}=p_i-\pi D^2/2$. In Fig.~\ref{fqhbraid}(b), we study the anyon statistics as a function of $D$. When $D$ is large enough, $(p_i)_{\textrm{anyon}}$ exponentially converges to the theoretical value $\pm\pi$. The critical value of $D$ for which $(p_i)_{\textrm{anyon}}$ is close enough to the theoretical value can be used as another definition of the quasihole size. If we set $|(p_i)_{\textrm{anyon}}+\pi|\leq0.01\pi$ as the threshold, we get the critical value of $D$ as $4.54\ell_B$, leading to $R_{\textrm{qh}}^{\textrm{FQH}}=D/2\approx2.27\ell_B$.

We emphasize that our results, although obtained for a square sample, are robust with respect to the aspect ratio and twist angle of the torus, so long as $|\textbf{L}_1|$ and $|\textbf{L}_2|$ are bigger than the quasihole size.

\section{Quasiholes in FCIs}
\label{sec:fci}
\begin{figure}
\centerline{\includegraphics[width=\linewidth]{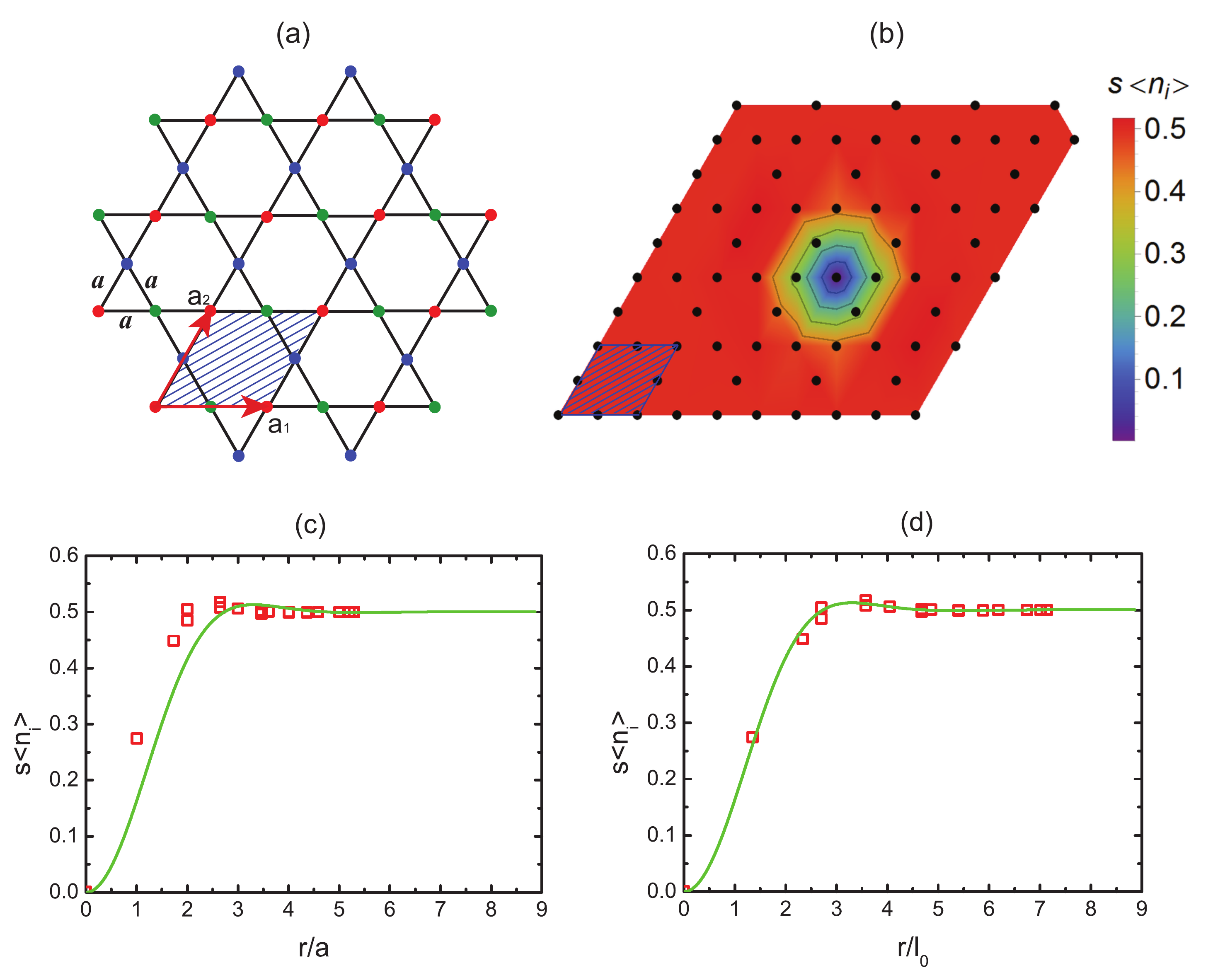}}
\caption{(Color online) (a) The kagome lattice with lattice vectors ${\bf a}_1$ and ${\bf a}_2$ and lattice constant $a$. Each unit cell, as indicated by the shade, contains $s=3$ sites (red, green and blue). (b) The lattice site occupation for the single-quasihole excitation of $\nu=1/2$ bosonic FCI in the kagome lattice model with $N=12,N_1\times N_2=5\times5$. The 2D distribution is obtained by fitting the discrete lattice site occupations by a continuous function (done by \emph{Mathematica} routine \textsc{LISTDENSITYPLOT}). (c)--(d) The lattice site occupation versus the distance from the quasihole in the FCI. The length units on the lattice are $a$ in (c) and $\ell_0$ in (d), respectively. The radial density distribution around the quasihole in the FQH case [Fig.~\ref{fqhden}(b)] is also plotted for reference (green line), with the length unit $\ell_B$.}
\label{kgden}
\end{figure}

Now we focus on fractional Chern insulators. We consider $N$ bosons on a 2D lattice on the torus with lattice constant $a$. There are $N_1$ and $N_2$ unit cells along two lattice vectors ${\bf a}_1$ and ${\bf a}_2$ respectively, and each unit cell contains $s$ sites [see. Fig.~\ref{kgden}(a) for an example based on the kagome lattice]. The single-particle problem is described by a tight-binding Hamiltonian $H_0=\sum_{{\bf k}} \sum_{\alpha,\beta=1}^s h_{\alpha\beta}({\bf k})d_{{\bf k},\alpha}^\dagger d_{{\bf k},\beta}$ with ${\bf k}$ in the first Brillouin zone, where $d_{{\bf k},\alpha}^\dagger (d_{{\bf k},\alpha})$ creates (annihilates) a boson with momentum ${\bf k}$ on the site $\alpha$ in a unit cell. The filling factor in one Bloch band is defined as $\nu\equiv N/(N_1N_2)$.
Bosons interact through the on-site two-body Hubbard Hamiltonian $H_{\textrm{int}}=\sum_i n_i(n_i-1)$, where $n_i$ is the occupation operator on lattice site $i$.
This interaction mimics the contact interaction of the FQH system in Sec.~\ref{sec:continuum} and stabilizes the $\nu=1/2$ bosonic FCIs in various lattice models.\cite{Kapit-PhysRevLett.105.215303,Wang-PhysRevLett.107.146803,Liu-PhysRevB.87.205136,Cooper-PhysRevLett.110.185301}
We use a simple on-site impurity $H_{\textrm{imp}}(w)=n_w$ to pin one quasihole on site $w$.
For large numerical efficiency, we project the interaction as well as impurity potentials into the Hilbert subspace of the lowest Bloch band that is fractionally occupied by bosons.
This projection is implemented by replacing $d_{{\bf k},\alpha}$ by $u_{\alpha,{\bf k}}\gamma_{{\bf k}}$, where $u_{\alpha,{\bf k}}$ is the $\alpha$ component of the eigenvector of the occupied band and $\gamma_{{\bf k}}$ is the annihilation operator of a particle with momentum $\bf {k}$ in the occupied band.

Similar to the FQH case, the FCI ground states of $H_{\textrm{int}}$ are (in general approximately) two-fold degenerate for $N_1N_2=2N$, and $N_{\textrm{qh}}$ delocalized quasiholes are created when the lattice size is enlarged to $N_1N_2=2N+N_{\textrm{qh}}$. These quasiholes are associated with a manifold of low-energy (generally not zero-energy) states in the energy spectrum of $H_{\textrm{int}}$. Again, we assume impurities do not mix the quasihole manifold with higher-energy
excited states. Therefore we can safely diagonalize impurity potentials in the quasihole manifold to obtain the ground states with localized quasiholes. The prediction of the counting of delocalized or localized quasiholes in FCIs can be obtained from the corresponding FQH results by the FQH-FCI mapping proposed in Ref.~\onlinecite{Bernevig-2012PhysRevB.85.075128}.

\subsection{Single quasihole}
\label{fcia}
We start our discussion by considering the kagome lattice model\cite{Tang-PhysRevLett.106.236802,Wu-PhysRevB.85.075116} as defined in Fig.~\ref{kgden}(a). In a manner similar to the FQH case, we first focus on the spatial extent of the quasihole.
The manifold with a single delocalized quasihole consists of $N_1N_2$ low-energy states. The diagonalization of the on-site impurity in this manifold gives us two approximately degenerate ground states. By computing the average lattice site occupation $\langle n_i\rangle$ over these two states, we find that the quasihole is indeed localized on one site where the occupation is very small. The lattice site occupation around the quasihole is not isotropic like the density in the FQH case, but it is inversion symmetric with respect to the quasihole, and $s\langle n_i\rangle$ tends to $\nu=1/2$ on sites far away from the quasihole [Fig.~\ref{kgden}(b)].

Because the FCIs in Chern number $|C|=1$ band can be regarded as the lattice version of corresponding FQH states, we expect to establish a correspondence between the quasihole spatial structure of these two systems. We plot the lattice site occupation versus the distance from the pinning potential. We compare it with the radial density distribution around the quasihole in the FQH case. If we naively use $a$ and $\ell_B$ as the length units in the FCI and FQH cases respectively, these two plots do not match as seen in Fig.~\ref{kgden}(c). However, there is no reason to expect that $a$ is the counterpart of $\ell_B$. In order to find the optimal length unit $\ell_0$ on the lattice that can mimic $\ell_B$ in the FQH case, we numerically minimize the following function
\begin{eqnarray}
f(\ell_0)=\sum_{i=1}^{sN_1N_2}|s\langle n(r_i)\rangle_{\textrm{FCI}}-2\pi\ell_B^2\rho_{\textrm{FQH}}(r_i\ell_0^{-1}\ell_B)|^2,
\label{densitydiff}
\end{eqnarray}
where $\langle n(r_i)\rangle_{\textrm{FCI}}$ is the occupation on lattice site $i$ with distance $r_i$ from the quasihole in FCIs with single-quasihole excitation, and $\rho_{\textrm{FQH}}(r)$ is the radial density distribution around the quasihole of corresponding FQH states. By using the FQH data $\rho(r)$ obtained in Sec.~\ref{continuuma} on a square torus (this is appropriate when the aspect ratio and twist angle of the lattice are not too small),
we get $\ell_0\approx0.742a$ for $N=7,N_1\times N_2=3\times5$ and $\ell_0\approx0.741a$ for $N=12,N_1\times N_2=5\times5$. With $\ell_0$ as the length unit on the lattice, the radial site occupation around the quasihole in the FCI indeed matches the curve in the FQH case [Fig.~\ref{kgden}(d)].

We interpret this numerically obtained $\ell_0$ as an approximation of the lattice magnetic length $\ell_B^{\textrm{lat}}$ of the underlying lattice. In the kagome lattice model, there is no net external magnetic field,\cite{Tang-PhysRevLett.106.236802} so it is unclear how to define a magnetic length as in the FQH case. However, we notice that the total phase that a particle picks up by hopping around the unit cell [indicated in Fig.~\ref{kgden}(a)] is $0$ and there is no observable difference between a phase $0$ and $2\pi$. Therefore, we can imagine that an effective magnetic field exists that causes an AB phase $2\pi$ for a boson moving around the unit cell. Then the lattice magnetic length can be defined as $\ell_B^{\textrm{lat}}\equiv\sqrt{A/(2\pi)}$, where $A$ is the area of the unit cell. For the kagome lattice model, $A=2\sqrt{3}a^2$, which immediately leads to $\ell_B^{\textrm{lat}}=(\sqrt{3}/\pi)^{\frac{1}{2}}a\approx0.743a$ that is very close to $\ell_0$.

In order to further examine the existence of the effective magnetic field in the unit cell, we use the quasihole as a detector. Because the charge of a quasihole is half of that of a boson, the AB phase that a quasihole picks up by moving around the unit cell should be $\pm\pi$ if the effective magnetic field really exists. We adopt the method in Ref.~\onlinecite{Kapit-PhysRevLett.108.066802} to move the quasihole by a time-dependent impurity $H_{\textrm{imp}}=\sum_j V_j(t)n_j$. At each time $t$, $V_j(t)$ is nonzero only on two nearest-neighbor sites $w_1$ and $w_2$ and $H_{\textrm{imp}}=(1-\eta)n_{w_1}+\eta n_{w_2}$. When $\eta$ slowly changes with $t$ from $0$ to $1$ (we use $150$ steps in the calculations), a quasihole is gradually moved from $w_1$ to $w_2$. Then similarly it can be moved from $w_2$ to the next site. We suppose the quasihole returns to the initial site at $t=T$.
Similar to Sec.~\ref{continuuma}, the unitary Berry matrix is
\begin{eqnarray}
\mathcal{B}=P\exp\Big\{-2\pi\textrm{i}\int_{0}^T\gamma(t) dt\Big\},
\label{berryfci}
\end{eqnarray}
where $\gamma_{ij}(t)=\textrm{i}\langle\psi_i(t)|\nabla_{t}|\psi_j(t)\rangle$ is the Berry connection matrix, $|\psi_i(t)\rangle$ are the approximately degenerate states that we get by diagonalizing the impurity potentials at each $t$, and $P$ is the time ordering symbol. By imposing a smooth gauge condition $\langle\psi_i(t)|\psi_j(t+dt)\rangle=\delta_{ij}+\mathcal{O}(dt^2)$, we have $\mathcal{B}_{ij}=\langle\psi_i(T)|\psi_j(0)\rangle$. The eigenvalues of $\mathcal{B}$ are $(e^{-\textrm{i}p_1},e^{-\textrm{i}p_2})$, where $p_1$ and $p_2$ are the AB phases that the quasihole picks up.
We calculate $(p_1,p_2)_{\textrm{AB}}$ for the kagome lattice model of different sizes. The results are very close to the expected value $\pm\pi$. We obtain $(p_1,p_2)_{\textrm{AB}}=(0.975\pi,-0.975\pi)$ and $(p_1,p_2)_{\textrm{AB}}=(0.998\pi,-0.998\pi)$ for $3\times5$ and $5\times5$ lattices, respectively.
This means that the quasihole indeed feels the effective magnetic field.

\begin{figure*}
\centerline{\includegraphics[width=\linewidth]{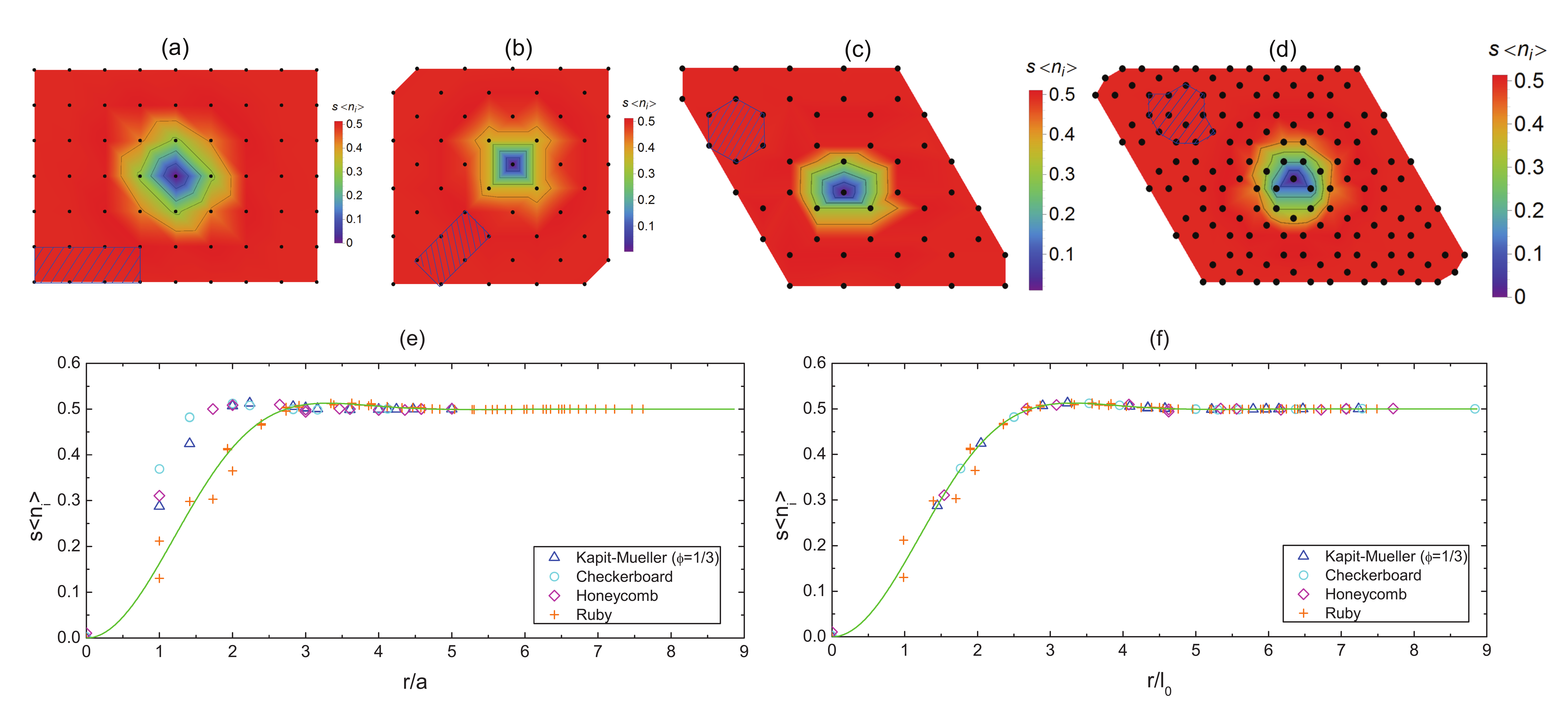}}
\caption{(Color online) (a)--(d) The 2D lattice site occupation for the single-quasihole excitation of $\nu=1/2$ bosonic FCI in the (a) Kapit-Mueller, (b) checkerboard, (c) honeycomb, and (d) ruby lattice model. The system size is $N=10,N_1\times N_2=3\times7,\phi=1/3$ in (a), and $N=12,N_1\times N_2=5\times5$ in (b)--(d). The lattice constant $a$ in each model is the distance between two nearest-neighbor sites. The unit cell in each lattice model is indicated by the shaded area, along which the AB phase that a quasihole picks up is expected to be $\pm\pi$. In (d) and (e) we show the radial lattice site occupation around the quasihole, where the length unit is $a$ in (d) and $\ell_0$ in (e). The radial density distribution around the quasihole in the FQH case [Fig.~\ref{fqhden}(b)] is also plotted for reference (green line) with the length unit $\ell_B$.
One can see that the radial site occupation around the quasihole in each FCI model is actually a discrete sampling of the density in the FQH case after we choose appropriate lattice length unit $\ell_0$ that depends on the model. }
\label{fciden}
\end{figure*}

\begin{table*}
\caption{\label{t1} Summary of our results of a single quasihole in various FCI models, including model parameters in the tight-binding Hamiltonian $H_0$, the lattice magnetic length $\ell_B^{\textrm{lat}}$, the numerical rescaling constant $\ell_0$ that minimizes Eq.~(\ref{densitydiff}), the quasihole radius $R_{\textrm{qh}}^{\textrm{FCI}}$, and the AB phase $(p_1,p_2)_{AB}$ that the quasihole obtains after moving around the unit cell. We adopt the same notation of model parameters as in Refs.~\onlinecite{regnault-PhysRevX.1.021014,Wu-PhysRevB.85.075116}. For the Kapit-Mueller model, we fix $N=10,N_1\times N_2=3\times7$, but consider different flux densities $\phi=1/2$ and $\phi=1/3$. For other FCI models, we consider two system sizes where $N=7,N_1\times N_2=3\times5$ and $N=12,N_1\times N_2=5\times5$ respectively. When calculating $\ell_0$, we use the FQH data $\rho(r)$ in Sec.~\ref{continuuma} on a square torus. This is appropriate because the aspect ratios and twist angles of our lattices are not too small. Besides $\ell_0$, we also give the theoretical values of $\ell_B^{\textrm{lat}}$ in brackets for comparison. $R_{\textrm{qh}}^{\textrm{FCI}}$ is quantified by Eq.~(\ref{rfci}) with $R_{\textrm{qh}}^{\textrm{FQH}}\approx1.76\ell_B$ and $2.27\ell_B$.}
\renewcommand{\multirowsetup}{\centering}
\begin{ruledtabular}
\begin{tabular}{cccccccc}
FCI models&parameters&$\ell_B^{\textrm{lat}}/a$&$\ell_0/a$&$R_{\textrm{qh}}^{\textrm{FCI}}/a$&$(p_1,p_2)_{\textrm{AB}}/\pi$\\
\hline
\multirow{2}{2cm}{kagome}&$t_1=1,$&
\multirow{2}{2cm}{$(\frac{\sqrt{3}}{\pi})^{\frac{1}{2}}$}&$3\times5:0.742$ $(0.743)$&
\multirow{2}{2cm}{$1.31\sim1.69$}&$3\times5:(0.975,-0.975)$\\
&$\lambda_1=0.9$& &$5\times5:0.741$ $(0.743)$&&$5\times5:(0.998,-0.998)$\\
\hline
\multirow{2}{2cm}{Kapit-Mueller}
&\multirow{2}{2cm}{N/A}&\multirow{2}{2cm}{$(\frac{1}{2\pi\phi})^{\frac{1}{2}}$}&$\phi=\frac{1}{2}:0.566$ $(0.564)$&$\phi=\frac{1}{2}:0.99\sim1.28$
&$\phi=\frac{1}{2}:(0.989,-0.989)$\\
& & & $\phi=\frac{1}{3}:0.691$ $(0.691)$&$\phi=\frac{1}{3}:1.22\sim1.57$&$\phi=\frac{1}{3}:(1.000,-1.000)$\\
\hline
\multirow{4}{2cm}{checkerboard}&$t_1=1,$&
\multirow{4}{2cm}{$(\frac{1}{\pi})^{\frac{1}{2}}$}&&&\\
&$t_2=0.5,$& &$3\times5:0.566$ $(0.564)$&
\multirow{2}{2cm}{$0.99\sim1.28$}&$3\times5:(0.999,-0.999)$\\
&$M=0,$& &$5\times5:0.566$ $(0.564)$&&$5\times5:(1.000,-1.000)$\\
&$\phi=\pi/4$& &&&\\
\hline
\multirow{4}{2cm}{honeycomb}&$t_1=1,$&
\multirow{4}{2cm}{$(\frac{3\sqrt{3}}{4\pi})^{\frac{1}{2}}$}&&&\\
&$t_2=1,$& &$3\times5:0.650$ $(0.643)$
&\multirow{2}{2cm}{$1.13\sim1.46$}&$3\times5:(0.923,-0.923)$\\
&$M=0,$& &$5\times5:0.648$ $(0.643)$&&$5\times5:(0.986,-0.986)$\\
&$\phi=0.2$& &&&\\
\hline
\multirow{5}{2cm}{ruby}&$t_r=1,$&
\multirow{5}{2cm}{$(\frac{3+2\sqrt{3}}{2\pi})^{\frac{1}{2}}$}&&&\\
&$t_i=1.2,$& &$3\times5:1.017$ $(1.014)$&
\multirow{3}{2cm}{$1.79\sim2.30$}&$3\times5:(0.999,-0.999)$\\
&$t_{1r}=-1.2,$& &&&\\
&$t_{1i}=2.4,$& &$5\times5:1.017$ $(1.014)$&&$5\times5:(1.000,-1.000)$\\
&$t_{4}=-1.46$& &&&
\end{tabular}
\end{ruledtabular}
\end{table*}

Our analysis above for the kagome lattice model also holds for other FCI models, such as the Kapit-Mueller,\cite{Kapit-PhysRevLett.105.215303,Liu-PhysRevB.88.205101} checkerboard,\cite{Sun-PhysRevLett.106.236803,neupert-PhysRevLett.106.236804,regnault-PhysRevX.1.021014} honeycomb,\cite{Haldane-PhysRevLett.61.2015,Wu-PhysRevB.85.075116} and ruby lattice model.\cite{Hu-PhysRevB.84.155116,Wu-PhysRevB.85.075116}
We summarize the results in Fig.~\ref{fciden} and Table \ref{t1}. For each model we consider two different system sizes to show the robustness of our results. In the Kapit-Mueller model, where a uniform external magnetic field $B$ exists, the magnetic length can be straightforwardly calculated by the standard definition as $\sqrt{\hbar/(eB)}=a(\frac{1}{2\pi\phi})^{\frac{1}{2}}$ with $\phi$ the magnetic flux density in each plaquette. Our definition of $\ell_B^{\textrm{lat}}$ leads to exactly the same result for $\phi=1/q (q=2,3,...)$, and the numerically obtained $\ell_0$ is also close to $a(\frac{1}{2\pi\phi})^{\frac{1}{2}}$. This fact strongly supports our definition of $\ell_B^{\textrm{lat}}$ in FCI models with zero net external magnetic field.

Now, we can establish a mapping between the FCI site occupation and the FQH density: the radial site occupation around the quasihole in a FCI model is actually a discrete sampling of the density of the corresponding FQH state, if the length units are $\ell_B^{\textrm{lat}}$ and $\ell_B$ in respective systems. On a finite lattice with fixed tight-binding parameters, $\ell_B^{\textrm{lat}}$ is approximated by $\ell_0$. With this mapping, we can define the quasihole radius in a FCI by
\begin{eqnarray}
R_{\textrm{qh}}^{\textrm{FCI}}=(R_{\textrm{qh}}^{\textrm{FQH}}/\ell_B)\ell_B^{\textrm{lat}}.
\label{rfci}
\end{eqnarray}

\begin{figure}
\centerline{\includegraphics[width=\linewidth]{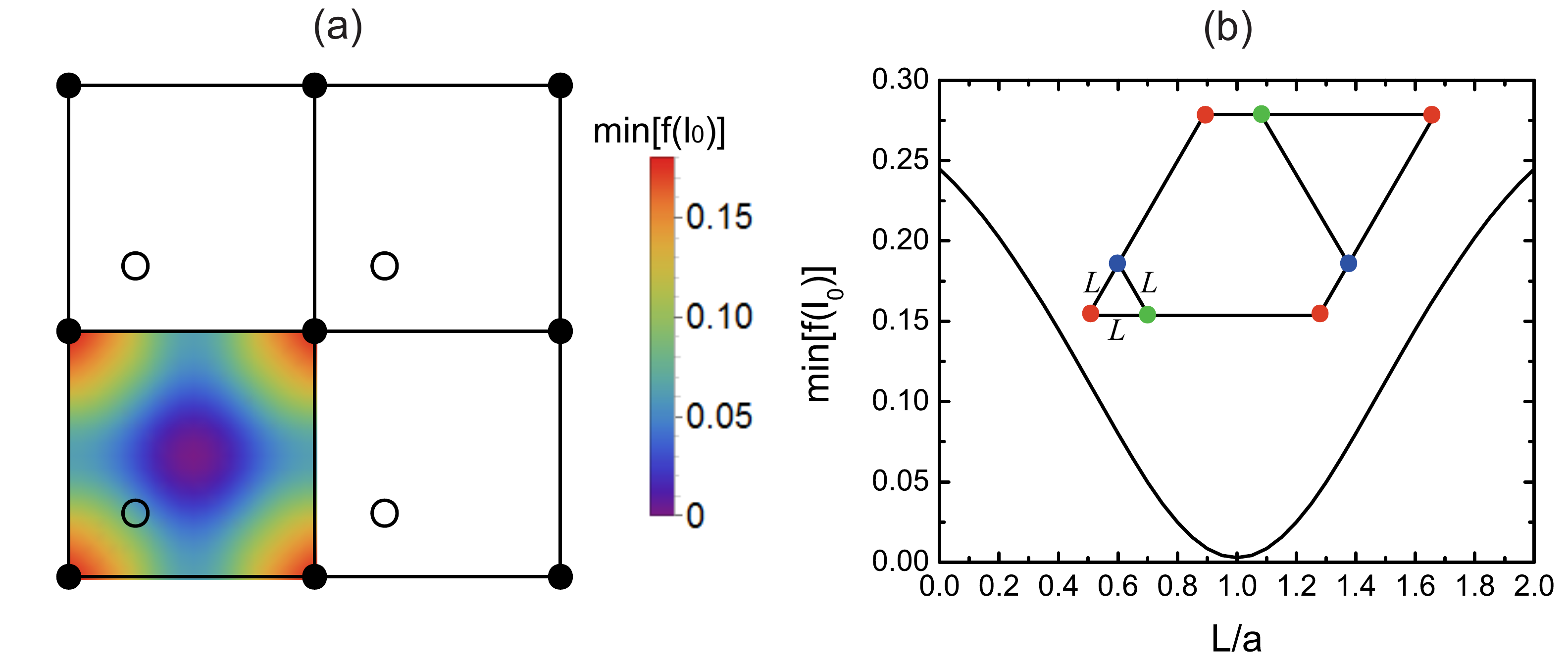}}
\caption{(Color online) The minimal value of $f(\ell_0)$ for different embedding of sites in a unit cell. (a) In each plaquette of the checkerboard lattice model, we fix the sites (solid dots) in corners, but change the position of the site (empty dot) in the plaquette. The minimal value of $f(\ell_0)$ almost goes to $0$ when the mobile site arrives at the center of the plaquette, which is the optimal embedding. (b) In the unit cell of the kagome lattice model, we keep three sites as an equilateral triangle with side length $L$. The minimal value of $f(\ell_0)$ has a valley at $L=a$, corresponding to the optimal embedding.}
\label{embedding}
\end{figure}

We end this section with a discussion of the effect of embedding on the FCI-FQH correspondence of the radial site occupation (density) around the quasihole. An embedding is determined by the positions of $s$ sites in one unit cell. Its information is not included in the tight-binding Hamiltonian, the interaction, or the impurity potential. However, it was previously shown that the embedding has to be carefully chosen to maximize the overlap between the numerical states and the model states \cite{Wu-PhysRevLett.110.106802} and establish the single-mode approximation.\cite{Repellin-PhysRevB.90.045114} In our calculation, the embedding does matter if we want to study the spatial extent of the quasihole because it determines the distance between lattice sites. For the kagome lattice model, the optimal embedding is the one shown in Fig.~\ref{kgden}(a).\cite{Repellin-PhysRevB.90.045114,Wu-PhysRevLett.110.106802} This embedding preserves the inversion symmetry which is also present in the interaction and energy spectrum. In this sense, we use the optimal embedding of each FCI model in the analysis above. If we change the embedding, we expect to see that the mapping between FCI site occupation and FQH density could become worse and even collapse. In Fig.~\ref{embedding}, we choose different embedding for the checkerboard and kagome lattice model. For each embedding, we compute the minimal value of $f(\ell_0)$ [Eq.~(\ref{densitydiff})]. Indeed, $\min[f(\ell_0)]$ is close to $0$ for the optimal embedding, and becomes much larger if the embedding is far from the optimal one. This means, the FCI-FQH mapping favors the optimal embedding rather than a random one.

\begin{figure}
\centerline{\includegraphics[width=\linewidth]{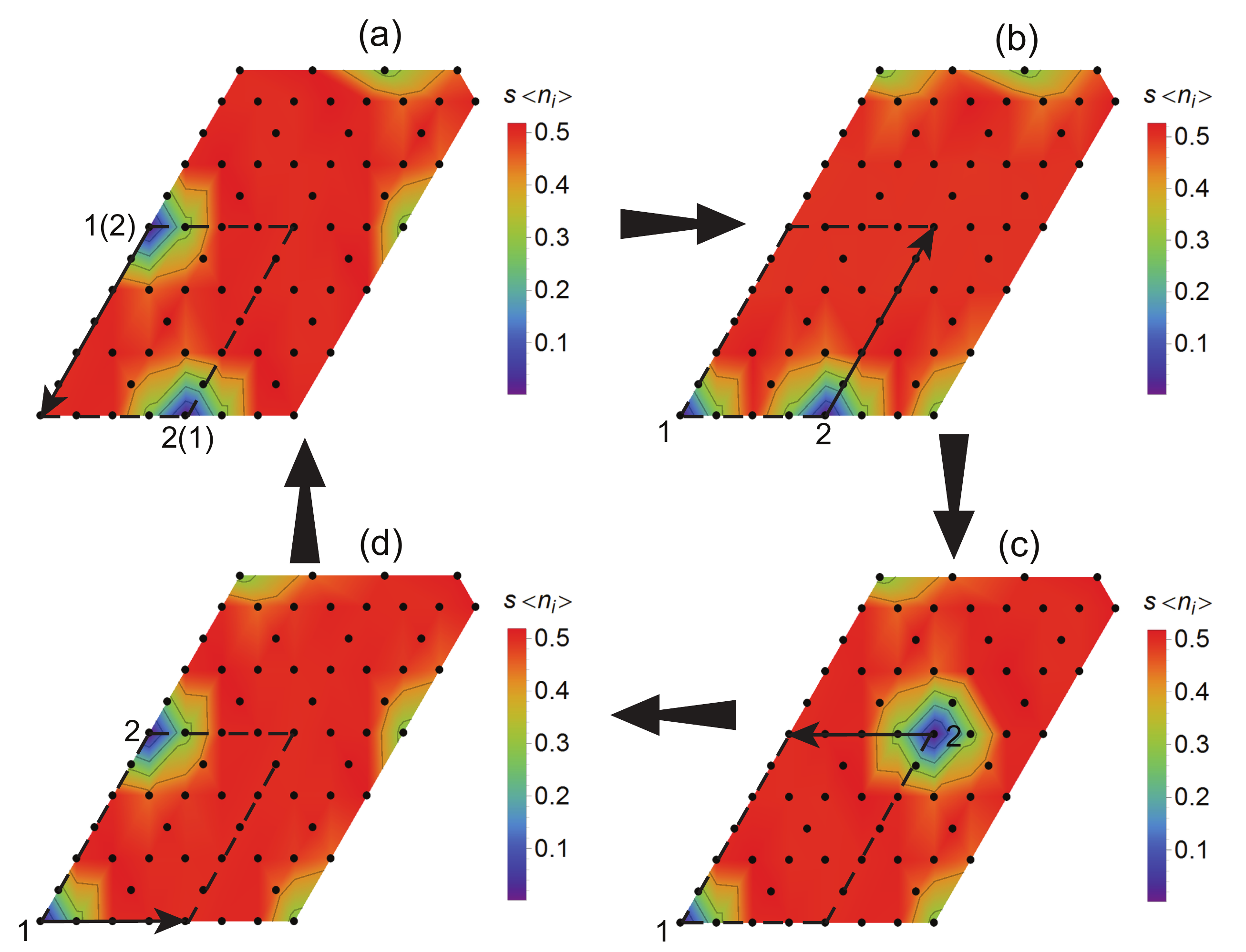}}
\caption{(Color online) (a)--(d) The path (dashed line) of exchanging two quasiholes on the $4\times6$ kagome lattice ($N=11$). This path encloses six unit cells. (a) In the initial configuration, quasihole $1$ is on the left and quasihole $2$ is at the bottom. Then quasihole $1$ moves downward but quasihole $2$ stays at the initial position. (b),(c) After quasihole $1$ arrives the lower left corner, quasihole $2$ first moves upward and then leftward. (d) When quasihole $2$ reaches the initial position of quasihole $1$, quasihole $1$ moves rightward to the initial position of quasihole $2$. This completes the exchange of two quasiholes, and the final positions of two quasiholes are indicated in the brackets in (a).}
\label{braid}
\end{figure}

\begin{table*}
\caption{\label{t2} Here we show the total Berry phases $(p_1,p_2)$ induced by the exchange of two quasiholes along the path in Figs.~\ref{braid}(a)--\ref{braid}(d), and the AB phases $(p_1,p_2)_{\textrm{AB}}$ that a single quasihole picks up along that path for various system sizes.}
\begin{ruledtabular}
\begin{tabular}{cccccccc}
system size $(N_{\textrm{qh}}=2)$&$(p_1,p_2)/\pi$&system size $(N_{\textrm{qh}}=1)$&$(p_1,p_2)_{\textrm{AB}}/\pi$\\
\hline
$N=7,N_1\times N_2=4\times4$&$(0.555,0.517)$&$N=7,N_1\times N_2=3\times5$&$(0.050,-0.050)$\\
$N=9,N_1\times N_2=4\times5$&$(0.511,0.490)$&$N=10,N_1\times N_2=3\times7$&$(0.050,-0.050)$\\
$N=11,N_1\times N_2=4\times6$&$(0.511,0.489)$&$N=12,N_1\times N_2=5\times5$&$(0.002,-0.002)$
\end{tabular}
\end{ruledtabular}
\end{table*}

\subsection{Braiding of two quasiholes}
\label{fcib}
We now address the braiding of quasiholes in FCIs. We generate two quasiholes on the lattice by letting $N_1N_2=2N+2$, so we can exchange them or move one around the other by a time-dependent impurity similar to the one used in Sec.~\ref{fcia}. At each time $t$, $V_j(t)$ is nonzero only on three sites $w_1$, $w_2$ and $w_3$. The impurity has the form of $(1-\eta)n_{w_1}+\eta n_{w_2}+n_{w_3}$, so it can pin one quasihole at $w_3$, and move the other from $w_1$ to $w_2$. Again, the total Berry phase $(p_1,p_2)$ can be split into two parts: one is the AB phase $(p_1,p_2)_{\textrm{AB}}$ caused by moving a single quasihole in the effective magnetic field along this path without other quasiholes enclosed; the other comes from the anyonic statistics $(p_1,p_2)_{\textrm{anyon}}$. We first exchange two quasiholes on the kagome lattice. The exchange path is shown in Figs.~\ref{braid}(a)--\ref{braid}(d). We choose this path to maximally separate two quasiholes during the whole process and enclose an integer number of unit cells. Thus we expect the total Berry phase $(p_1,p_2)=(p_1,p_2)_{\textrm{anyon}}=(0.5\pi,0.5\pi)$ and the AB phase $(p_1,p_2)_{\textrm{AB}}=(0,0)$. Our numerical results are indeed close to the expected values (Table \ref{t2}). One should note that they are improved for larger system size and more isotropic aspect ratio.

\begin{figure}
\centerline{\includegraphics[width=0.8\linewidth]{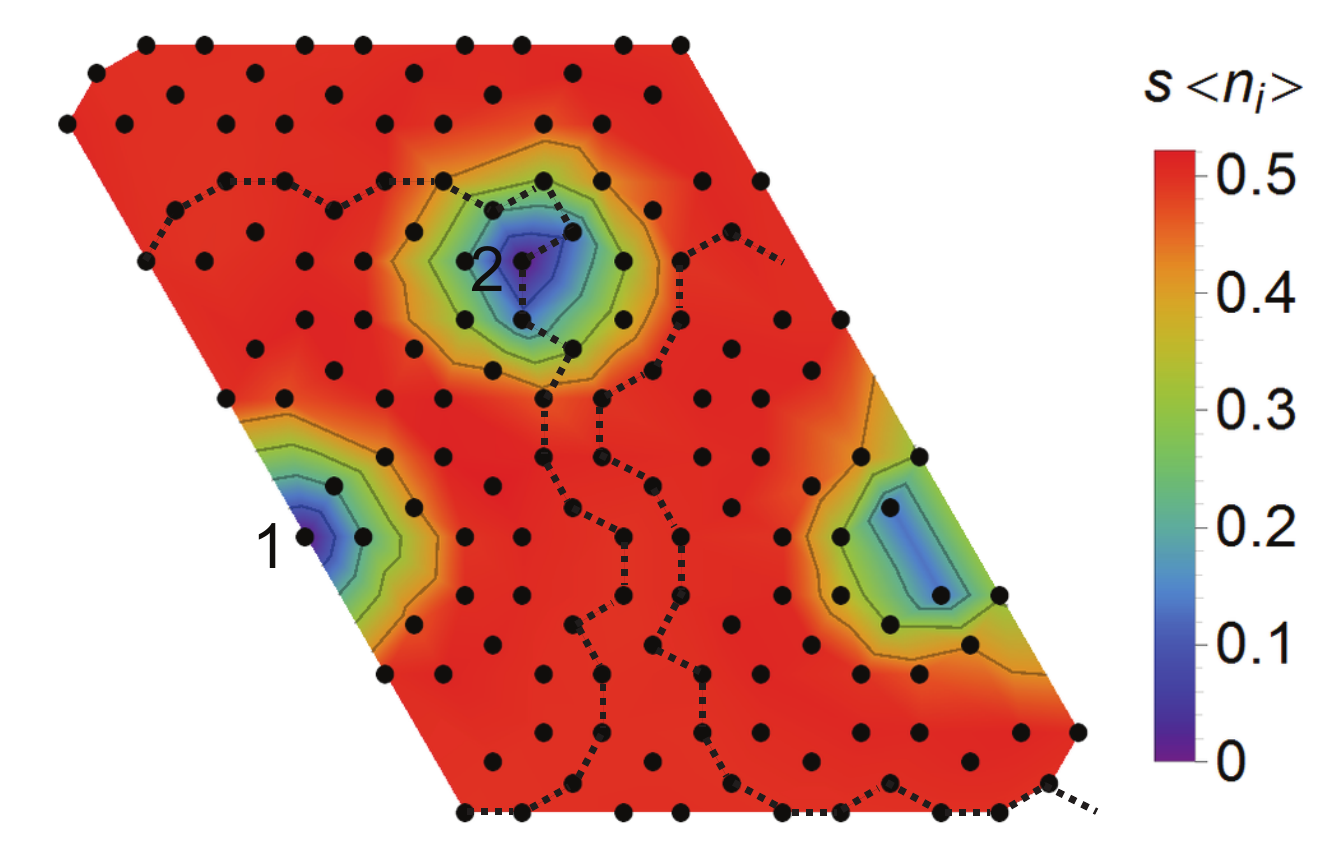}}
\caption{(Color online) The path (dashed line) of moving one quasihole around the other on the $4\times6$ ruby lattice ($N=11$). Quasihole 1 is fixed one one site, and quasihole 2 moves clockwise along the path for one loop. Contrary to Fig.~\ref{braid}, this path does not enclose an integer number of unit cells, producing an additional AB phase when braiding two quasiholes.}
\label{braid2}
\end{figure}

We can of course choose a path that does not enclose an integer number of unit cells. For example, we can braid one quasihole around the other along a complicated path on the ruby lattice  as shown in Fig.~\ref{braid2}. We compute the AB phase by moving a single quasihole along the same path. The numerical results are $(p_1,p_2)=(0.219\pi,0.220\pi)$ (obtained on the $4\times6$ lattice) and  $(p_1,p_2)_{\textrm{AB}}=(-0.778\pi,-0.778\pi)$ (obtained on the $5\times5$ lattice), therefore $(p_1,p_2)_{\textrm{anyon}}=(0.997\pi,0.998\pi)$, which is extremely close to the expected value $(p_1,p_2)_{\textrm{anyon}}=(\pi,\pi)$.

\section{Conclusion}
\label{sec:conclusion}

In this paper, we have investigated the properties of quasiholes for the $\nu=1/2$ Abelian bosonic FQH and FCI states. For the model Laughlin states on the torus geometry, we pin the $-e/2$ quasihole by a $\delta$ impurity and measure the quasihole radius $R_{\textrm{qh}}$ through the second moment of the radial density distribution and the braiding phase of two quasiholes. From these two we find $R_{\textrm{qh}}\approx1.76\ell_B$ and $R_{\textrm{qh}}\approx2.27\ell_B$ respectively.

Similarly, on the lattice, a FCI quasihole is pinned by an on-site impurity. The determination of the FCI quasihole radius is a bit more involved due to the natural anisotropy of the lattice. However, by rescaling the discrete set of on-site occupation around a FCI quasihole to fit the corresponding FQH quasihole density profile, we are able to reliably extract the FCI quasihole spatial extent for all models considered here. The seemingly model-dependent quasihole radius can be unified once it is expressed in unit of the model-dependent lattice magnetic length $\ell_B^{\textrm{lat}}$: it is then very close to $R_{\textrm{qh}}$ in unit of $\ell_B$.

We also perform the braiding of two quasiholes and we accurately recover the predicted statistical phase. Interestingly, the braiding path can be chosen to enclose an integer number of unit cells and thus to cancel any Aharonov-Bohm contribution to the Berry matrix, providing a direct access to the statistical phase.

While our work focuses on the simplest FQH/FCI state, namely the Laughlin $\nu=1/2$ phase, we argue that the relation between the sizes of any quasihole in the FQH effect and its counterpart in any FCI model is related by the model-dependent lattice magnetic length $\ell_B^{\textrm{lat}}$. Knowing many quasihole types have been considered and studied in the context of the FQH effect, this relation is highly valuable for any experimental detection scheme that relies on the spatial extent of the excitations.

There are several possible future developments based on this work. Checking the non-Abelian statistics in FCIs is more challenging. Indeed, the tractable lattice sizes by exact diagonalization are smaller and the quasiholes are larger \cite{Toke-PhysRevLett.98.036806,Wu-PhysRevLett.113.116801} (and thus harder to separate). Moreover realistic pinning potentials are more complicated.\cite{Prodan-PhysRevB.80.115121,Storni-PhysRevB.83.195306} Therefore, DMRG-based techniques are more appropriate for studying the non-Abelian problem on the lattice.\cite{Liu-PhysRevB.88.081106,Grushin-2014arXiv1407.6985G} Secondly, we have only used isotropic short-range interactions in this work. It might also be useful to consider the effect of long-range interaction such as the dipolar interaction,\cite{Yao-PhysRevLett.110.185302,Liu-PhysRevB.88.205101,Hafezi:2007p67} which may increase the quasihole size. The anisotropic interaction may distort the quasihole and reveal the quantum geometry in FCIs.\cite{PhysRevB.88.115117,PhysRevLett.107.116801} Another interesting route is the FCIs for band with a Chern number $|C|$ larger than one.\cite{max,Wang-PhysRevB.86.201101,Yang-PhysRevB.86.241112,Liu-PhysRevLett.109.186805,Sterdyniak-PhysRevB.87.205137,Wu-PhysRevLett.110.106802,Wu-2013arXiv1309.1698W,Sterdyniak-2014arXiv1410.0357S} Once strong interactions are turned on in one $|C|>1$ band,  these systems host new phases that are a generalization of the Halperin states \cite{Halperin83} in the boundary conditions that entangle the orbital and internal degrees of freedom.\cite{Wu-PhysRevLett.110.106802} Looking at the effect of this interplay on the braiding properties might be instructive in probing these phases.

\begin{acknowledgments}
We thank Emil Bergholtz, Sonika Johri, Yang-Le Wu and Duncan Haldane for discussions. Z.~L. and R.~N.~B. were supported by the Department of Energy, Office of Basic Energy Sciences through Grant No.~DE-SC0002140. N.~R. was supported by ANR-12-BS04-0002-02, the Princeton Global Scholarship, DARPA SPAWARSYSCEN Pacific N66001-11-1-4110, MURI-130-6082, Packard Foundation, and Keck grant.
\end{acknowledgments}

\bibliography{FCI_hole}

\end{document}